\documentclass[final,12pt]{elsarticle}

\usepackage{graphicx}
\usepackage{amssymb}
\usepackage{color}
\biboptions{sort&compress}

\journal{Journal of Molecular Spectroscopy}

\begin{document}

\begin{frontmatter}

\title{Broadband analysis techniques for Herschel/HIFI spectral surveys of chemically rich star-forming regions\tnoteref{1}}
\tnotetext[1]{Herschel is an ESA space observatory with science instruments provided by European-led Principal Investigator consortia and with important participation from NASA.}

\author[label1]{Justin L. Neill}
\ead{jneill@umich.edu}
\author[label1]{Edwin A. Bergin}
\author[label2]{Dariusz C. Lis}
\author[label2]{Thomas G. Phillips}
\author[label2]{Martin Emprechtinger}
\author[label3]{Peter Schilke}

\address[label1]{Department of Astronomy, 500 Church St., Ann Arbor, MI 48109, USA}
\address[label2]{California Institute of Technology, Cahill Center for Astronomy and Astrophysics, 301-17, Pasadena, CA 91125, USA}
\address[label3]{I. Physikalisches Institut, Universit\"{a}t zu K\"{o}ln, Z\"{u}lpicher Str. 77, 50937 K\"{o}ln, Germany}

\begin{abstract}
The Heterodyne Instrument for the Far Infrared (HIFI) aboard the Herschel Space Observatory has acquired high-resolution broadband molecular spectra of star-forming regions in a wavelength range that is mostly inaccessible from ground-based astronomical observatories.  These spectral surveys provide new insight into the chemical composition and physical properties of molecular clouds.  In this manuscript, we present initial results from the HIFI spectral survey of the Sagittarius B2(N) molecular cloud, which contains spectral features assigned to at least 40 different molecules in a range of physical environments.  While extensive line blending is observed due to the chemical complexity of this region, reliable molecular line identifications can be made, down to the noise floor, due to the large number of transitions detected for each species in the 1.2 THz survey bandwidth.  This allows for the extraction of new weakly emitting species from the line forest.  These HIFI surveys will be an invaluable archival resource for future investigations into interstellar chemistry.
\end{abstract}


\end{frontmatter}

\section{Introduction}

Broadband molecular line surveys are an effective tool for investigating the rich molecular inventory of star-forming regions\cite{Herbst2009, Turner1991, Nummelin2000, Belloche2008, Belloche2009, Remijan}.  From detailed analyses of the spectral signatures of molecular clouds, their chemical abundances and physical properties (e.g. temperature and density) can be determined.  The derived parameters can then be used to test and guide models of the source and develop our understanding of the physics and chemistry of the interstellar medium.  These surveys are often made available to the public, which allow for other researchers to pursue more detailed investigations or to search for new molecular species in the data.
 
A number of new radio astronomical observatories, both single-dish telescopes and interferometeric arrays, have recently begun operations that offer significant advances in bandwidth and sensitivity over older facilities.  These new observatories will lead to a dramatic increase in the number of publicly available spectral surveys in the coming years.  Developing strategies to aid in the analysis and interpretation of these spectra, therefore, is an important goal in the field of astrochemistry.  The large number of data channels associated with these surveys necessitates efficient methods for modeling and visualization of the data and results, as the spectrum should be fit globally---characterizing all molecules simultaneously, rather than one at a time, and across the full survey bandwidth.  Millimeter and submillimeter spectra of chemically rich regions are characterized by a dense line forest at lower signal levels, made up largely of features from complex organics, which emit at many frequencies due to their large partition functions.  Many features therefore consist of blends that contain contributions from several molecules.  Once the spectral signatures of known molecules are characterized down to the noise limit, new species, which can be hidden in the spectrum by the features of molecules that emit more strongly, can be identified.

This manuscript presents early results from a spectral survey toward Sagittarius B2(N) acquired using the Heterodyne Instrument for the Far Infrared (HIFI)\cite{deGraauw2010} aboard the Herschel Space Observatory\cite{Pilbratt2010}.  Sgr B2(N) has the greatest observed chemical complexity of any molecular cloud in our galaxy, and a large fraction of the $\sim$170 molecules detected thus far in the interstellar medium\cite{Woon} were first detected in this source.  This region has been the target of a number of broadband spectral surveys using ground-based observatories in the centimeter and millimeter wavelength regions\cite{Turner1991, Nummelin2000, Belloche2008, Belloche2009, Remijan}.  HIFI is a high-resolution spectrometer with continuous coverage from 480--1250 GHz and 1410--1910 GHz.  Most of this frequency range is inaccessible from the ground due to atmospheric absorption, so HIFI offers a new look into a well-studied molecular source.  Additionally, by providing a census of which molecules are emitting in each frequency range, this survey can be used to aid in characterizing the submillimeter spectra of other molecular clouds.  We present an analysis of the emission spectrum of methanol in this spectrum, which is detected through a large number of transitions ranging widely in frequency and excitation energy. This data set serves as an example of the Herschel legacy spectral surveys that we expect will be highly useful to the astronomical community in the coming years.

\section{Observations and modeling approach}

The observations reported here were acquired as part of the Herschel Observations of EXtra-Ordinary Sources (HEXOS) guaranteed time key program\cite{Bergin2010}.  The HEXOS program consists primarily of complete HIFI surveys of five sources within the Sagittarius B2 and Orion star-forming regions:  Sgr B2(N), Sgr B2(M), Orion KL, Orion S, and the Orion Bar.  The Sgr B2(N) observations presented in this manuscript (HIFI bands 2b, 3a, and 3b) were acquired on 16--17 September 2010 using the wide band spectrometer (WBS) with a spectral resolution of 1.1 MHz, pointed towards the Sgr B2(N) hot core with coordinates $\alpha_{\mathrm{J}2000} = 17^{\mathrm{h}}47^{\mathrm{m}}19^{\mathrm{s}}.88$ and $\delta_{\mathrm{J}2000} = -28^{\circ}22'18''.4$.  HIFI is a double sideband (DSB) spectrometer, and each frequency was observed with a redundancy of 8 (i.e., with eight different local oscillator settings) in order to eliminate ghost transitions in the spectrum.  Data reduction was performed using version 8.0 of the Herschel Interactive Processing Environment (HIPE)\cite{Ott2010}.  After standard pipeline processing and spur removal, the continuum was fit and removed.  Strong lines (with line minus continum antenna temperature greater than 10 K) were removed and flagged, and the \emph{doDeconvolution} task in HIPE was performed on the data sets both with and without the strong lines included.  The strong lines were then added back into the deconvolved single-sideband (SSB) spectra, along with the fit continuum.  Spectra were obtained in both instrumental polarizations, which in the figures presented here were averaged together to improve the signal-to-noise ratio.  No channel weighting or gain correction were included in the data reduction.  All of the figures presented here are of deconvolved SSB spectra, smoothed to a velocity resolution of $\sim$0.7 km/s.

The deconvolved spectra were modeled using XCLASS\cite{Schilke}, which accesses the CDMS\cite{Muller2005} and JPL\cite{Pickett1998} spectral databases.  This software makes the local thermodynamic equilibrium (LTE) approximation, which assumes that all of the energy level populations for a given spatial component can be described by a Boltzmann distribution.  This approach is a necessary first step for large-bandwidth spectral analyses such as these; a wide variety of programs that go beyond the LTE approximation are available, but are significantly more computationally intensive to implement.  Additionally, for many of the molecules observed in this spectrum, especially the more complex molecules (with ``complex" here being defined as having six or more atoms\cite{Herbst2009}) that make up a large fraction of the line density, rates for collisional excitation by H$_2$, necessary to solve for statistical equilibrium between radiative and collisional processes, either have not been calculated, or the rates available in the literature do not cover the energy or temperature range observed in surveys (especially for sub-mm/far-IR surveys such as with HIFI).  The LTE approximation is a good one in the limit of dense gas, as is seen in the Sgr B2(N) hot core ($n($H$_2) > 10^7$ cm$^{-3}$)\cite{Qin2011}.

\section{Results and discussion}

In Figure 1 we show a portion of the HIFI spectrum toward Sgr B2(N), demonstrating the molecular diversity seen in this source.  Panel (A) shows the spectrum from 714--801 GHz (band 2b of HIFI), which represents 7\% of the total spectral coverage.  Earth's atmosphere is opaque in most of this band due to the water transition at 752 GHz.  A large number of features are observed, both in emission and in absorption against the source continuum, which in this frequency region arises from thermal emission from dust.  The molecular carriers of the strongest features are identified.  Methanol is the only complex organic to have strong features ($|\Delta T_{mb}| > 1$ K) in this band; this molecule is observed to have high line density in many surveys of dense molecular clouds due to its high abundance and complex asymmetric top spectrum, which make it a common tool with which to constrain the physical structure of dense molecular clouds\cite{Leurini2007, Wang2010, Kama2010}.

\begin{figure*}
\centering
\includegraphics[width=5.5 in]{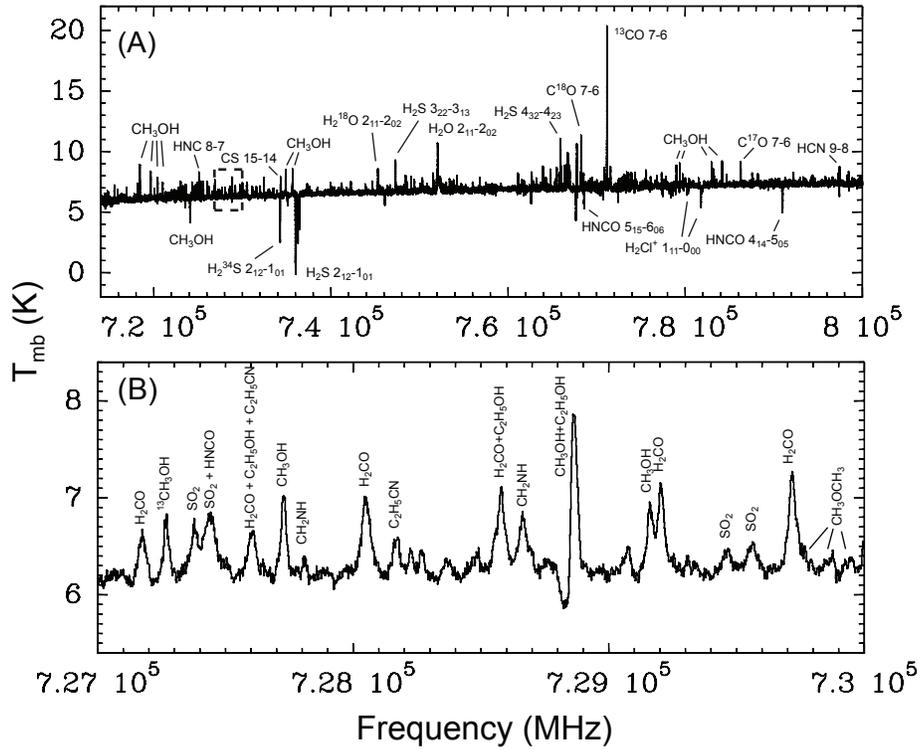}
\caption{Panel A: The submillimeter spectrum toward Sgr B2(N) from 714-801 GHz as measured by HIFI.  Panel B: An expanded view of the spectrum from 727-730 GHz (the dashed box in panel A).  In both panels, the assigned molecular carriers of the strongest lines based on full-band modeling are labeled.  The observed spectra have been corrected to an LSR velocity of 64 km s$^{-1}$, the primary velocity component of the Sgr B2(N) molecular core.}
\end{figure*}

Panel B shows a 3 GHz subset of this spectrum (the region indicated with a dashed box in panel A), with molecule assignments based on our preliminary modeling labeled.  It can be seen that, in addition to methanol, seven other molecules (including complex species like dimethyl ether, ethanol, and ethyl cyanide) have features detected in this frequency range.  Where more than one molecule has been listed in the assignment for a particular feature, our model calculates that both (or all) molecules have features that contribute substantially to the flux of the observed feature.  Each of the species identified in this figure has been detected through a large number of transitions in this survey.  To date we have identified approximately 40 molecules, which range in size from light hydrides (e.g. water, ammonia) to complex molecules such as methanol, dimethyl ether (CH$_3$OCH$_3$), and methyl formate (HCOOCH$_3$).  Most molecular features are attributed to the Sgr B2(N) core; the observed central velocities and linewidths vary from molecule to molecule due to different spatial distributions and excitation, but for most molecules the dominant component has a central velocity of +64 km/s and a linewidth of $\sim$7-10 km/s, as in previous analyses of this source\cite{Turner1991,Nummelin2000,Belloche2008,Belloche2009}.  Additionally, some simpler molecules are found to have features arising from line-of-sight clouds with different velocity shifts, as have been previously measured toward this source\cite{Wirstrom2010,Polehampton2005,Neufeld2003}.  The full analysis of this spectrum is currently in progress, and will be presented in a future publication.

While most of the strongest features in this survey are readily identified, making confident line identifications of weaker transitions is challenging due to the combination of baseline excursions (which are common at the $\sim$0.1 K level) and line blending.  Many of the unidentified lines are likely due to transitions of molecules that are included in the full-band model, but are currently not in the spectral catalogs.  Laboratory work to identify these transitions, which are due to vibrationally excited states, isotopologues, or other weak features, is needed, and is being undertaken for some abundant complex species\cite{Fortman2010}.  Nevertheless, particularly for complex species with high line densities, it is possible to make confident identifications of catalogued transitions down to the noise floor, if a large number of transitions from each molecule are observed with consistent kinematic parameters and intensities based on modeling, and no lines predicted by the model to be present are missing.  In recent years several new weakly emitting complex molecules have been identified in Sgr B2(N) at millimeter wavelengths with this approach\cite{Belloche2008,Belloche2009}.

Figure 2 shows an example of the LTE modeling procedure and weak line identification using the spectrum of methanol.  Panel A shows a strong Q-branch which has been used to fit the parameters of the LTE model, particularly rotational temperature and column density.  A key advantage to bands like this one is that a large number of lines are found in a small frequency range, so that frequency-dependent effects such as beam dilution and dust absorption do not affect the relative intensities between transitions.  As the transitions span a wide range in lower-state energy, from 90.9 cm$^{-1}$ for the $J_K = 5_5 - 5_4$ transition at highest frequency in the plot (835003.87 MHz) to 511.5 cm$^{-1}$ for the $J_K = 23_5 - 23_4$ transition at lowest frequency (830720.59 MHz), this band allows for the rotational temperature to be well constrained.  The best-fit LTE model, which was refined using the full HIFI spectrum of methanol, is overplotted.  This model predicts that $\sim$2000 lines of methanol present in the spectral catalogs have intensities greater than 0.2 K (including transitions from $^{12}$CH$_3$OH, $v_t = 0,1,2$ and $^{13}$CH$_3$OH, $v_t = 0,1$).

Two components are required to achieve a satisfactory fit to the methanol emission spectrum, the parameters of which are presented in Table 1; we note that while this model is fairly successful in modeling the emission of methanol across the broad-bandwidth HIFI spectrum, this fit should still be regarded as preliminary as the full-band analysis is still ongoing.  The warmer component in this fit is optically thick in the stronger lines, while the colder component is optically thin and contributes emission only to the lowest-energy lines.  We also have included dust extinction in this analysis, deriving a dust optical depth of 1.1 at 835 GHz (assuming a spectral index of 2).  All of the parameters shown in Table 1, with the exception of the source size of the colder component (which is not well constrained), were fitted as free parameters in the model.  Both components have a central LSR velocity of 64 km/s and a width of 9 km/s.

\begin{figure*}
\centering
\includegraphics[width=4.8 in]{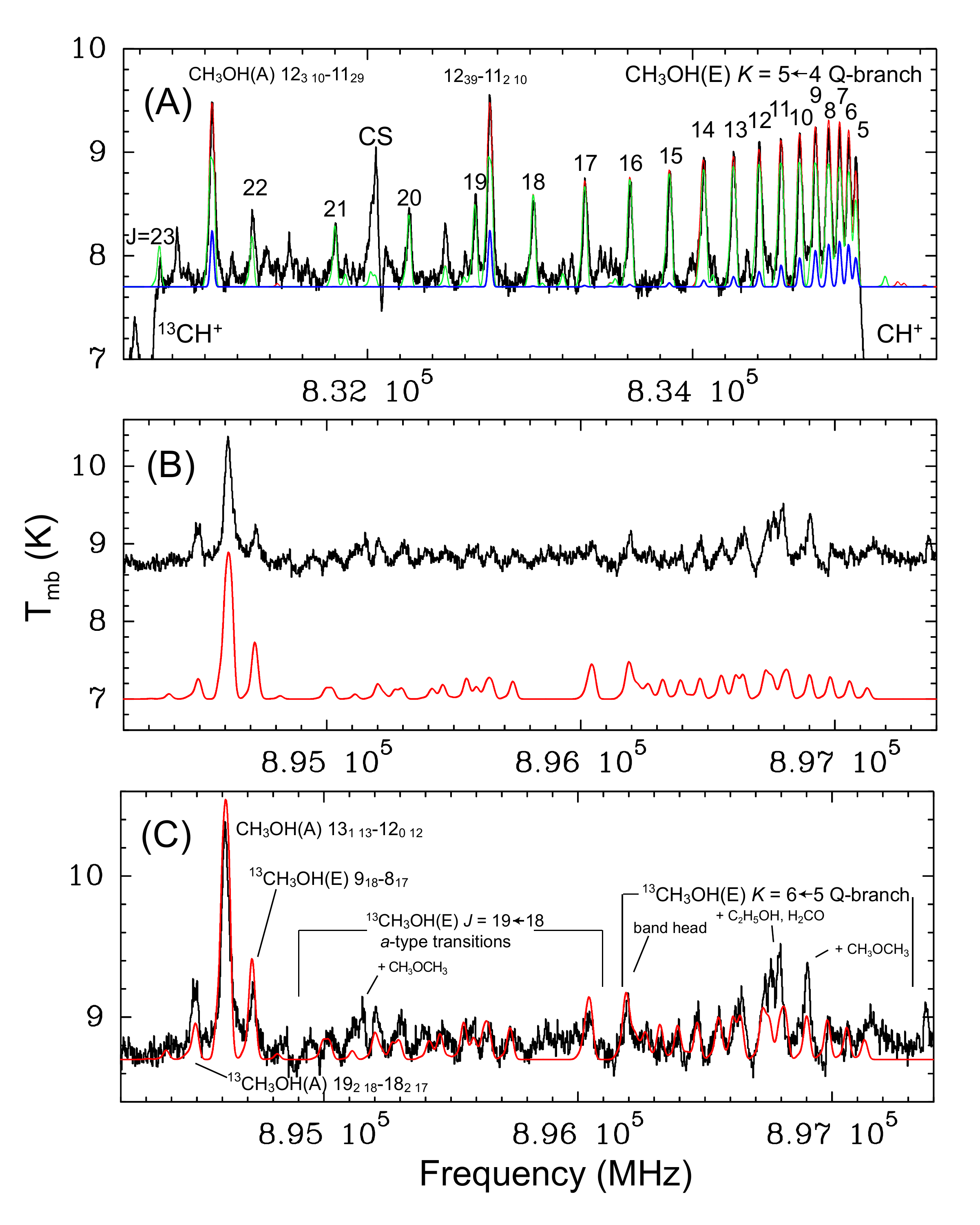}
\caption{Illustration of the modeling of weak lines of methanol in the HIFI spectrum of Sgr B2(N).  In each panel, the red trace is the best-fit LTE model to the spectrum of methanol as described in the text.  In panel A, the contributions of the individual components are plotted, with the warmer (170 K) component shown in green and the cooler (60 K) shown in blue.  The observed spectra have been corrected to an LSR velocity of 64 km s$^{-1}$, the primary velocity component of the Sgr B2(N) molecular core.} 
\end{figure*}

\begin{table}
\begin{center}
\caption{Fit LTE parameters of methanol in Sgr B2(N).}
\begin{tabular}{l c c c}
\hline
& Source size ($''$) & $T_\mathrm{rot}$ (K) & $N_\mathrm{tot}$ (cm$^{-2}$) \\
\hline
Warmer component & 3.5 & 170 & $4.0 \times 10^{18}$ \\
Cooler component & 10 & 60 & $6.0 \times 10^{16}$ \\
\hline
\end{tabular}
\end{center}
\end{table}

Panel B shows a different portion of the spectrum which also contains a number of lines of methanol, primarly of the $^{13}$CH$_3$OH isotopologue.  In this model, we have assumed the standard Galctic Center $^{12}$C/$^{13}$C ratio of 20, based on observations of a variety of molecular species in Sgr B2 and nearby clouds\cite{Wilson1994, Muller2008, Milam2005}.  Panel C shows the observed and model spectra overlaid, demonstrating that much of the structure observed in the frequency range is in fact well reproduced by the LTE model spectrum.  This figure should serve to emphasize that even features barely above the noise floor are attributable to molecular carriers and can be reliably modeled when a large number of features are observed for a given molecule.  Many of the weak features of methanol and other complex organics that clutter the spectrum can be identified through this process, which will enable searches for other molecules among the many residual lines.

\section{Conclusion}

We anticipate that the Herschel/HIFI spectral surveys collected as part of the HEXOS key program will be a valuable legacy for the molecular astrophysics community.  The survey described here of the Sagittarius B2(N) molecular cloud, in particular, covers a largely unexplored frequency range and provides a near-complete chemical inventory of the most chemically complex source in our galaxy with high sensitivity and spectral resolution.  The large bandwidth of the survey allows for the detection of a large number of lines for each molecular species, which imposes rigorous constraints on the physical parameters of the region.  Additionally, the detailed inventory of which molecules emit in each frequency range in the most chemically rich source in the Galaxy will be a valuable legacy for observers characterizing the submillimeter spectra of other molecular sources.

The approach described in this manuscript points the way forward to methods for the comprehensive analysis of rich spectra.  Unlike previous broadband surveys of Sgr B2(N) and other star-forming regions, where a single band is characterized, the analysis presented here involves modeling the entire 1.2 THz HIFI spectrum consistently.  This is greatly helped by the fact that this spectrum was obtained in a uniform way and on the same instrument, with high calibration and pointing accuracy and without weather variations from band to band due to the stable environment of space.  The frequency coverage of HIFI overlaps with the highest-frequency bands of the ground-based Atacama Large Millimeter/Submillimeter Array (ALMA), which has the potential of measuring spectra in every atmospheric window between 30 and 900 GHz.  The techniques described here for modeling high bandwidth interstellar spectra will therefore be applicable to that very sensitive instrument.  Other facilities will be available in the near future that can measure large bandwidth spectra:  the Stratospheric Observatory for Infrared Astronomy (SOFIA) will have broadband spectral coverage in the far-IR, and existing ground-based millimeter facilities such as the Institut de Radioastronomie Millim\'{e}trique (IRAM) 30 m telescope are implementing broadband receivers with high spectral resolution, so wide-bandwidth chemical surveys of molecular clouds are becoming increasingly routine.

We also expect that further insight into the physical and chemical conditions of star-forming regions from spectral surveys such as this will be aided by further laboratory and theoretical work from the chemical physics community.  The laboratory spectra of a large number of molecules of interstellar interest have not been recorded, or are incomplete, in the wavelength range of the HIFI instrument, and the rates of a number of relevant molecular processes (such as chemical reactions and collisional excitation) are not well known.  Developments on these fronts will have a critical role in the revolutionary science anticipated from this new generation of radio astronomical instrumentation.

\section{Acknowledgements}

HIFI has been designed and built by a consortium of institutes and university departments from across Europe, Canada, and the United States under the leadership of SRON Netherlands Institute for Space Research, Groningen, The Netherlands and with major contributions from Germany, France, and the US.  Consortium members are: Canada: CSA, U. Waterloo; France: CESR, LAB, LERMA, IRAM; Germany: KOSMA, MPIfR, MPS; Ireland: NUI Maynooth; Italy: ASI, IFSI-INAF, Osservatorio Astrofisico di Arcetri-INAF; Netherlands: SRON, TUD; Poland: CAMK, CBK; Spain: Observatorio Astron\'{o}mico Nacional (IGN), Centro de Astrobiolog\'{i}a (CSIC-INTA); Sweden: Chalmers University of Technology--MC2, RSS \& GARD, Onsala Space Observatory, Swedish National Space Board, Stockholm Observatory; Switzerland: ETH Zurich, FHNW; USA: Caltech, JPL, NHSC.  Support for this work was provided by NASA through an award issued by JPL/Caltech.


\end{document}